\documentclass[
aps,
reprint,
superscriptaddress,
amsmath,amssymb,
pra
]{revtex4-2}
\usepackage{hyperref}
\usepackage{graphicx}
\usepackage{dcolumn}
\usepackage{bm}

\begin{document}

\title{Entangling two levitated particles in free space via trap modulation and Bayesian feedback}

\author{Jinke Cao}
\affiliation{Department of Physics, Huazhong Normal University, Wuhan 430079, China}

\author{Qi Guo}
\affiliation{%
  State Key Laboratory of Quantum Optics and Quantum Optics Devices, Shanxi University, Taiyuan, Shanxi 030006, China
}
\affiliation{Collaborative Innovation Center of Extreme Optics, Shanxi University, Taiyuan 030006, China}

\author{Huatang Tan}
\email{tht@mail.ccnu.edu.cn}  
\affiliation{Department of Physics, Huazhong Normal University, Wuhan 430079, China}

\begin{abstract}
We propose an optimal control scheme for generating quantum entanglement between two optically-levitated nanoparticles in free space. Specifically, we consider that the mechanical motion frequencies of the two levitated particles are modulated by adjusting the amplitude of the trapping beam. The two particles are coupled through Coulomb interaction, and the particles' positions are continuously monitored via homodyne detection on the back-scattered light from both particles. By employing an optimal Bayesian feedback scheme, we achieve unconditional entanglement between the two particles in steady states. More precisely, a Kalman filter is used to estimate the states of the two particles and subsequently a linear quadratic regulator is applied to derive the optimal feedback forces exerted on the particles. Physically, periodic modulation enables significant quantum squeezing in both the common mode and the differential mode of the two particles. The Coulomb coupling between the particles introduces a difference in the squeezing of the two normal modes, thereby facilitating the entanglement of the two levitated particles. Our scheme allows for the realization of both conditional and unconditional entanglement at relatively low measurement efficiencies and with low requirements for Coulomb coupling strength, significantly enhancing the feasibility of implementation.
\end{abstract}

\maketitle

\section{INTRODUCTION}
Quantum entanglement \cite{RevModPhys.81.865}, as one of the core features of quantum mechanics. Its essence lies in the inseparability of subsystem states and non-local coordinations within quantum systems. Since the proposal of the Einstein-Podolsky-Rosen (EPR) paradox, research on entangled states has evolved from being a contentious issue in fundamental physics to becoming a cornerstone of quantum information technology \cite{RevModPhys.77.513}, finding widespread applications in fields such as quantum computing, quantum communication, and precision measurement. In recent years, there has been significant interest in macroscopic quantum entanglement, which not only aids in further elucidating the transition from the classical to the quantum world \cite{PhysRevLett.93.190402, PhysRevLett.97.237201, poot2012mechanical} but also provides invaluable resources for quantum sensing and quantum metrology \cite{barzanjeh2022optomechanics, RevModPhys.90.025004}. 

Cavity optomechanical \cite{RevModPhys.86.1391} systems serve as an excellent platform for realizing macroscopic quantum entanglement, with successive achievements in optomechanical entanglement \cite{PhysRevLett.98.030405, PhysRevA.78.032316, PhysRevA.84.052327} and mechanical-mechanical entanglement \cite{PhysRevA.87.022318, PhysRevA.89.063805, ockeloen2018stabilized, Li_2015}. Recently, levitated nanoparticles \cite{millen2020optomechanics, chang2010cavity, PhysRevA.83.013803, gonzalez2021levitodynamics} have emerged as a significant research focus, integrating findings from disciplines such as optomechanics, atomic physics, and control theory, and providing an ideal platform for both fundamental and applied research. Compared with traditional optomechanical systems, optically levitated nano-mechanical system significantly reduce contact with the thermal environment and eliminate dissipation caused by clamping. This enables levitated nanoparticles to exhibit longer coherence times, laying the foundation for coherent manipulation of mesoscopic mechanical systems and the generation of quantum entanglement between spatially separated systems. Experimentally, thermal squeezing \cite{PhysRevLett.117.273601} of trapped nanoparticles and ponderomotive squeezing \cite{PhysRevLett.129.053602} of light fields have been observed using rapid frequency switching of trapped nanoparticles and quantum measurement techniques, respectively. Very recently, quantum squeezing of the mechanical motion of levitated nanoparticles has been achieved in an optical lattice system by employing rapid transitions between frequencies and quantum measurement techniques \cite{2504.17944}. Phonon lasers based on levitated particles have also been observed by utilizing active devices and nonlinear effects \cite{kuang2023nonlinear}. In addition, advanced cooling \cite{PhysRevLett.109.103603, doi:10.1073/pnas.1309167110, asenbaum2013cavity, PhysRevLett.122.223601} and control \cite{PhysRevLett.121.033603, PhysRevLett.121.033602, PhysRevLett.117.123604, Kuhn:17, magrini2021real} technologies have created the conditions necessary for studying the quantum properties of levitated particles. For example, recent experiment has achieved optimal quantum control and cooling of levitated particles in free space via measurement and Beyessian  feedback \cite{tebbenjohanns2021quantum}. A scheme has also been proposed for cooling two levitated nanoparticles to their ground states using sideband cooling via optomechanical coupling \cite{PhysRevA.109.053521}. Recent theoretical study has also investigated steady-state unconditional entanglement between two levitated particles under strong Coulomb coupling and Markovian feedback \cite{PhysRevLett.129.193602}. This scheme requires a relatively high Coulomb coupling strength, which should be comparable to the mechanical frequency, making its experimental realization challenging. By employing an optimal feedback control scheme, unconditional entanglement can be achieved under weaker repulsive interactions \cite{2408.07492}. 

Periodic modulation is a common approach for investigating quantum behaviors in systems. In cavity optomechanical systems, this method is frequently employed to generate quantum entanglement and quantum squeezing \cite{bothner2020cavity}. Currently, by leveraging this technique, precise control over nonlinear stochastic bistable dynamics has been achieved in levitated particle systems \cite{ricci2017optically}. In our scheme, we investigate quantum entanglement in a levitated particle system by combining the linear quadratic Gaussian (LQG) control strategy with periodic modulation of the mechanical frequency. Periodic modulation of the mechanical frequency can significantly enhance quantum squeezing in the two mechanical normal modes (common mode and differential mode) \cite{PhysRevResearch.2.013052, PhysRevResearch.7.L012071}. The Coulomb interaction between levitated particles induces discrepancies in the squeezing amplitude or direction of quantum squeezing in these two normal modes, thereby generating entanglement between the two levitated particles. Homodyne detection reflects the positional information of the two particles while ensuring the stability of the system.  The states of the two particles are then estimated by a Kalman filter, and two feedback signals are generated via LQR control to minimize the excess noise, thereby achieving unconditional entanglement of the system. Our research results indicate that frequency modulation can significantly enhance both conditional and unconditional entanglement between two levitated particles, while substantially relaxing the parameter constraints for entanglement generation, providing a reference for the experimental preparation of unconditional entanglement. 

This paper is organized as follows. In Sec. II, we derived both the conditional and unconditional dynamical evolution equations for the system. In Sec. III, we investigated the conditional and unconditional entanglement of the system, as well as the mechanism of entanglement generation. Finally, we summarize our result in Sec. IV.
\begin{figure}[htb]
\includegraphics[height=9cm,width=7cm]{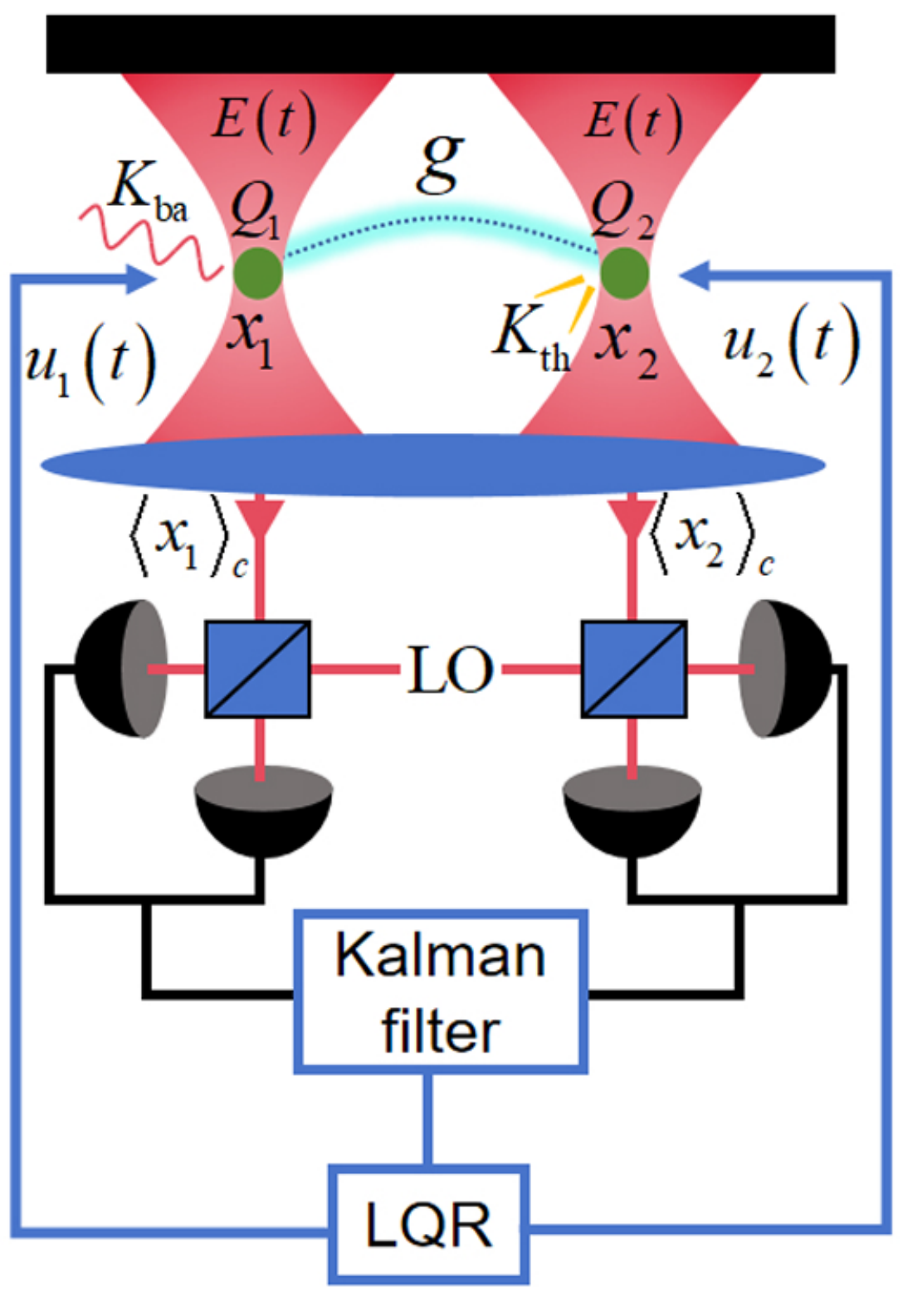}
\caption{\label{fig_model}Schematic diagram of a scheme for entangling two optically-levitated particles. Two particles that are coupled through Coulomb interaction (coupling constant $g$) are trapped in two optical traps, and the amplitude $E(t)$ of the light beam is periodically modulated. The random motion of gas molecules within the system can induce thermal decoherence (decoherence rate is $K_{\text{th}}$) on the particles.
Scattered photons introduce back-action noise characterized by a decoherence rate of $K_\text{ba}$, and these photons, carrying information about the particle's position, are detected by homodyne detection with a local oscillator (LO). A Kalman filter estimates the state of the two particles from the detection records. A LQR is employed to determine optimal feedback signals $u_1(t)$ and $u_2(t)$, which are applied to the two particles, thereby forming a feedback loop.}
\end{figure}
\section{MODEL AND SYSTEM DYNAMICS}
The system we are considering is depicted in Fig.~\ref{fig_model}. It consists of two particles, each with a mass of $m$, located at positions $X_{1,2}$ , with an average separation distance of $d$. The particles are trapped by two amplitude-modulated optical traps. The modulation amplitude of the traps is $E_t = E[{ 1+ \alpha \cos ({\Omega }t})]$, where $\Omega$ represents the modulation frequency and $\alpha$ denotes the modulation depth. Since the frequency of the particle's mechanical motion is related to the amplitude of the trapping optical field as $\omega _x(t) \propto E_t^2(t)$, the mechanical frequency of the particle can be expressed as ${{\omega }_{x}}\left( t \right)={{\omega }_{m}}{{\left[ 1+\alpha \cos \left( {\Omega}t \right) \right]}^{2}}$, where $\omega_m$ is the unmodulated frequency.

The Coulombic coupling interaction between the two particles is denoted as $H_\text{int} = Q_1Q_2/4\pi \varepsilon _0r$, where $r = \sqrt{(X_1 - X_2)^2 + d}$ represents the actual distance separating the particles and $Q_{1,2}$ signifies the charges carried by the particles. For small mechanical displacements, i.e., $X_{1,2} \ll d$, the interaction Hamiltonian  $H_\text{int}$ can be expanded up to the second order:
\begin{align}
   {\hat{H}_{\operatorname{int}}} = \hbar g{{\left( {\hat{x}_{1}}-{\hat{x}_{2}} \right)}^{2}}, 
\end{align}
where we have omitted the constant term since it does not affect the dynamical evolution of the system. Introducing the dimensionless position quadrature $\hat x_{k = 1,2} = X_k/x_\text{zpf}$ with the zero-point fluctuation $x_\text{zpf} = \sqrt{\hbar/m\omega_m}$, $g = - \frac{Q_1Q_2x_\text{zpf}^2}{8\pi\varepsilon_0\hbar d^3}$ is Coulombic coupling rate. By applying a time-varying electric field to the particles, the motion of the two particles can be controlled, and the resulting feedback Hamiltonian can be expressed as $H_\text{fb}^k = -Q_k E_\text{fb}^k(t) x_\text{zpf}\hat x_k = \hbar u_k(t)\hat x_k$, where $u_k(t)$ is the feedback force applied to the $k$-th particle. The Hamiltonian of the modulation system can be written as 
\begin{align}
  \hat{H}/{\hbar }\;  = & \sum\limits_{k =  1,2}{\left[ \frac{{{\omega }_{m}}}{2}\hat p_{k}^{2}+\frac{{{\omega }_{x}}\left( t \right)}{2}\hat x_{k}^{2} \right]} \nonumber  \\
  & +\sum\limits_{k =  1,2}{{{u}_{k}}\left( t \right){\hat{x}_{k}}}+g{{\left( {\hat{x}_{1}}-{\hat{x}_{2}} \right)}^{2}}, 
   \label{hmt1}
\end{align} 
where $\hat p_k$ is the dimensionless momentum operator for the mechanical motion of the $k$-th levitated particle. 

The scattered light carrying the positional information of the two particles is measured by two independent homodyne detectors, with the measurement results expressed as photocurrents $I_k(t)$
\begin{align}
{{I}_{k}}\left( t \right)=\sqrt{\eta {{K }_{\text{ba}}}}\left\langle {\hat{x}_{k}} \right\rangle_c dt+d{{W}_{k}}\left( t \right).
\end{align}
The term $K_\text{ba}\propto \omega_x(t)$ represents the decoherence rate introduced by scattered photons and the back-scattered light is measured by two independent homodyne detectors with a measurement efficiency $\eta$. $dW_k$ is the Wiener increment, a normally distributed variable that satisfies the properties $\mathbb{E} [dW_k]=0,\mathbb{E} [dW_idW_j]=\delta _{ij}dt$. The corresponding stochastic master equation is given by
\begin{align}
  d{\hat{\rho }_{c}}&=-\frac{i}{\hbar}\left[\hat H,{\hat{\rho }_{c}} \right]dt-\frac{i\gamma }{2}\sum\limits_{k=1,2}{\left[ {\hat{x}_{k}},\left\{ {\hat{p}_{k}},{{\hat \rho }_{c}} \right\} \right]}dt \nonumber  \\
 &-\frac{1}{2}\left( {{K }_{\text{th}}}+{{K }_{\text{ba}}} \right)\sum\limits_{k=1,2}{\left[ {\hat{x}_{k}},\left[ {\hat{x}_{k}},{{\hat \rho }_{c}} \right] \right]}dt \nonumber  \\
 &+\sqrt{\eta {{K }_{\text{ba}}}}\sum\limits_{k=1,2}{\left\{ {\hat{x}_{k}}-{{\left\langle {\hat{x}_{k}} \right\rangle }_{\text{c}}},{{\hat \rho }_{c}} \right\}d{{W}_{k}}},
\end{align}
where $\gamma$ represents the damping of two identical mechanical systems, and the scattering of residual gas in the system can induce thermal decoherence on the two particles, with a decoherence rate given by $K_\text{th}=\gamma \bar{n}$, where $\bar{n} \simeq k_BT/\hbar\omega_m$ represents the number of mechanical thermal phonons in the high-temperature limit (e.g., $T = 300~\mathrm K$ as adopted in this paper). The terms in last line characterize the nonlinear detection-induced backaction effects, and when the ensemble average is performed, these terms disappear.


We define the operator vector $\mathbf{X}=[\hat x_1, \hat p_1, \hat x_2, \hat p_2]^T$, the conditional mean $\left \langle \mathbf{X}  \right \rangle _c=\mathrm{Tr} [\mathbf{X}\hat \rho _c]$, and the covariance matrix $(V_c)_{ij}=\mathrm{Re}(\langle \mathbf{X}_i\mathbf{X}_j \rangle) - \langle \mathbf{X}_i\mathbf{X}_j\rangle$. The equations of motion for the conditional mean $\langle\mathbf{X}\rangle_c$ and covariance matrix $V_c$ can be uniformly expressed as \cite{PhysRevResearch.1.033161, PhysRevA.100.023843}
\begin{subequations}
\begin{align}
   d{{\left\langle \mathbf{X} \right\rangle }_{c}}& = \mathbf A{{\left\langle \mathbf{X} \right\rangle }_{c}}dt + \mathbf {Bu} + 2{{V }_{c}}\mathbf Cd\mathbf{w}, \label{e5a}\\ 
  \frac{d{{V }_{c}}}{dt}& = \mathbf A{{V }_{c}} + {{V }_{c}}{\mathbf{A}^{T}} + \mathbf N-4{{V }_{c}}\mathbf C{\mathbf{C}^{T}}{{V }_{c}}.
  \label{e5b}
\end{align}
\label{e5}
\end{subequations}
where $\mathbf B$ is feedback matrix, which indicates the degree of control over the two particles, and $\mathbf u(t) = [u_1(t), u_2(t)]^T$ is time-dependent feedback signal. The other matrices are
\begin{align}
\mathbf{A} & = \left[ \begin{matrix}
   0 & {{\omega }_{m}} & 0 & 0  \\
   -{{\omega }_{x}}\left( t \right) - 2g & -\gamma & 0 & 0  \\
   0 & 0 & 0 & {{\omega }_{m}}  \\
   0 & 0 & -{{\omega }_{x}}\left( t \right)-2g & -\gamma  \\
\end{matrix} \right], 
\nonumber \\
\mathbf{C} & = \sqrt{\eta {{K }_{\text{ba}}}}{{\left[ \begin{matrix}
   1 & 0 & 0 & 0  \\
   0 & 0 & 1 & 0  \\
\end{matrix} \right]}^{T}},
\nonumber \\
\mathbf N&=(K_{\text{ba}}+K_{\text{th}})\text{diag}(0,1,0,1).
\end{align}
We note that Eqs.(\ref{e5a}) and (\ref{e5b}) are formally equivalent to a classical Kalman filter \cite{wiseman2009quantum}.


We see from Eq.(\ref{e5a}) that the first moments are related to the measurement results and thus stochastic. On the contrary, the covariance matrix $V_c$ is independent of the outcomes and deterministic. The effect of continuous homodyne measurement is embodied by the last nonlinear term of Eq.(\ref{e5b}). By averaging over all possible outcomes of the measurement, one can obtain the unconditional state of the system $\hat \rho_u =\mathbb{E}[\hat \rho_c]$. The corresponding unconditional covariance matrix is
 \begin{align}
{{V }_{u}}={{V }_{c}}+{{V }_{\text{ex}}},
 \end{align}
where ${V }_{\text{ex}}=\mathbb{E}[\langle \mathbf{X}\rangle_c \langle \mathbf{X}\rangle_c^T]$ represents the excess noise matrix originating from the random walk of $\langle \textbf{X}\rangle_c$ in phase space and ensemble average. To mitigate the impact of excessive noise, a feedback strategy can be utilized. To determine the optimal feedback forces $u_k(t)$, a classical control theory can be used \cite{PhysRevA.60.2700}. Specifically, a Kalman filter can be employed as the optimal observer, and  a linear quadratic regulator as the optimal controller, with the objective of minimizing the following cost function
\begin{align}
j = \int_{{t_0}}^{t_1 }{dt\left( \mathbb{E} \left[ {{\mathbf{X} }_{c}}^{T}\mathbf{P}{{\mathbf{X}  }_{c}}+\mathbf{u} ^{T}\mathbf{Q}\mathbf{u} \right] \right)}
\end{align}
where $\mathbf{P}$ and $\mathbf{Q}$ are semi-definite cost matrix and positive-definite control effort matrix, respectively. The optimal feedback strategy can be obtained by solving the backward Riccati equation
\begin{align}
\frac{d\mathbf \Sigma}{d\left( -t \right)}={\mathbf{A}^{T}} \mathbf \Sigma+ \mathbf \Sigma\mathbf A+\mathbf{P}-\mathbf \Sigma \mathbf B{{\mathbf{Q}}^{-1}}{\mathbf{B}^{T}}\mathbf \Sigma,
\end{align}
the optimal control matrix $\mathbf K_\text{opt} = \mathbf Q^{-1}\mathbf B^T \mathbf \Sigma$, and the optimal feedback signal 
\begin{align}
\mathbf u(t) = - \mathbf K_\text{opt} \langle \mathbf{X}_c\rangle. 
\end{align}
Time-dependent feedback signal relies on the continuous estimation of the system's state. This strategy of making real-time adjustments to the feedback signal based on new information, which is analogous to classical Bayesian reasoning, is referred to as Bayesian feedback \cite{PhysRevA.66.013807}.  This is different from  Wiseman-Milburn feedback scheme \cite{wiseman2009quantum} in which the measurement sign is directly used drive the system. After establishing the optimal feedback protocol, the time evolution of the first-order moment vector mean $\langle\mathbf{X}_c\rangle$ and the excess noise matrix $V_\mathrm{ex}$ is given by:
\begin{align}
   d{{\left\langle \mathbf X \right\rangle }_{c}}  = & \left[\mathbf A-\mathbf{BK}_\text{opt} \right]{{\left\langle \mathbf X \right\rangle }_{c}}dt+2{{V }_{c}} \mathbf Cd\mathbf w, \nonumber \\ 
  \frac{d{{V }_\text{ex}}}{dt}  =& \left[\mathbf A-\mathbf{BK}_\text{opt} \right]{{V }_\text{ex}}+{{V }_\text{ex}}{{\left[\mathbf A-\mathbf{BK}_\text{opt} \right]}^{T}} \nonumber  \\ 
  &-4{{V }_{c}}\mathbf C{\mathbf{C}^{T}}{{V }_{c}}.
\end{align}

For unconditional states, the selection of feedback strategies will have a significant impact on unconditional entanglement. In the experimental implementation of the system described in this protocol, two primary feedback mechanisms are adopted: the identical feedback mechanism and the independent feedback mechanism. The feedback forces applied to two charged particles can achieve either identical control or independent control by employing a time-varying electric field of the same kind \cite{doi:10.1126/science.abp9941, Deplano:24, PhysRevResearch.5.013070, Bykov:23} or by arranging an array of optical traps \cite{vijayan2023scalable, reisenbauer2024non, PhysRevApplied.20.024018}, respectively. When identical feedback control is applied to both particles, to ensure the system's controllability, the two particles must carry unequal electric charges. Under this condition, the feedback matrices for the two feedback strategies are respectively given by
\begin{align}
  & {\mathbf{B}_{\text{ide}}}={{\left[ \begin{matrix}
   0 & 1 & 0 & \frac{Q_1}{Q_2}
\end{matrix} \right]}^{T}}, \nonumber  \\ 
 & {\mathbf{B}_{\text{ind}}}={{\left[ \begin{matrix}
   0 & 1 & 0 & 0  \\
   0 & 0 & 0 & 1  
\end{matrix} \right]}^{T}}.
\end{align}
The control effort matrices corresponding to these two scenarios are denoted as $\mathbf Q_\text{ide}=q/\omega_m$ and $\mathbf Q_\text{ind}= \text{diag}(q/\omega_m, q/\omega_m)$. Considering that we are only interested in the steady oscillatory entanglement between particles after a long period of evolution, we select a cost matrix $\mathbf P= \text{diag}(\omega_m, \omega_m, \omega_m, \omega_m)$ targeted at cooling both bosonic modes.
\begin{figure}[htb]
\includegraphics[height=9cm,width=9cm]{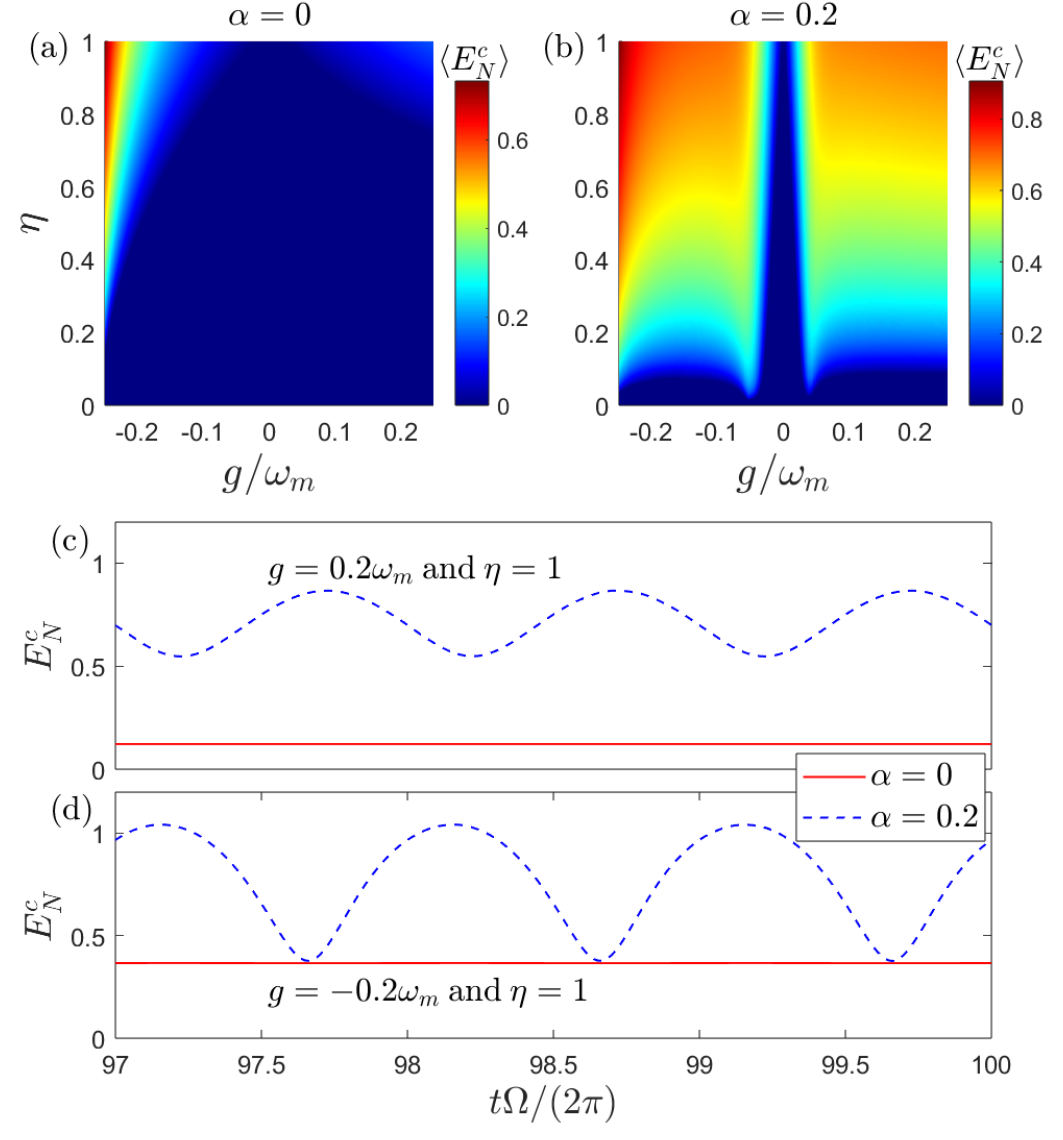}
\caption{\label{fig1}Time-averaged conditional entanglement $\langle E_N^c \rangle$ versus the Coulomb coupling rate $g$ and measurement efficiency $\eta$ for (a) no modulation $\alpha=0$ and (b) modulation $\alpha=0.2$. (c) and (d) represent the time evolution of the conditional entanglement $E_N^c$ under attractive ($g=0.2$) and repulsive ($g=-0.2$) interactions, respectively, at the optimal detection ($\eta=1$). Other parameters are $K_\text{ba}/\omega_x(t)=0.05$, $K_\text{th}/\omega_m=2.5\times 10^{-3}$, $\gamma/\omega_m=10^{-10}$ and $\Omega = 2 \omega_m$.}
\end{figure}
\section{THE GENERATION AND ENHANCEMENT OF ENTANGLEMENT}
We adopt the logarithmic negativity to quantify conditional (unconditional) entanglement $E_N^{c(u)}$, defined as 
\begin{align}
E_N^{c(u)} & = -\mathrm{ln} (2\tilde {v}_{c(u)}),
\end{align}
where $\tilde {v}_{c(u)}$ is the minimal symplectic eigenvalue of covariance matrix $V_c$ or $V_u$. 
\begin{figure}
\includegraphics[height=8cm,width=9cm]{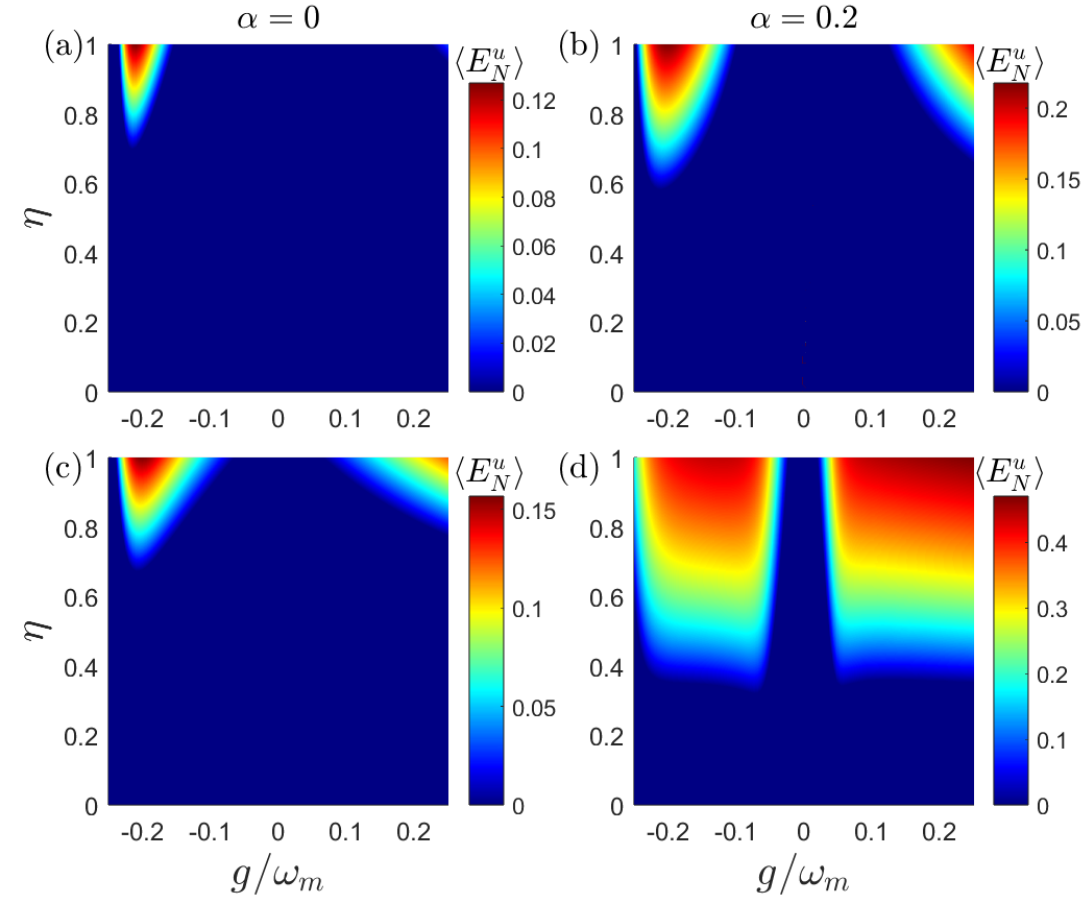}
\includegraphics[height=6cm,width=8.5cm]{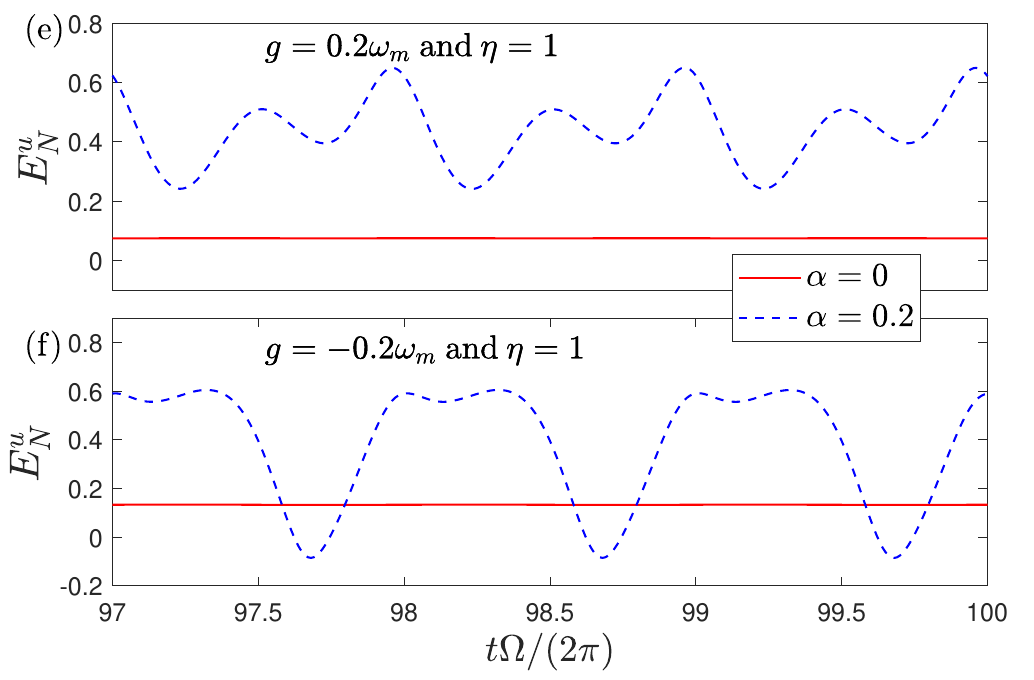}
\caption{\label{fig2}Time-averaged Conditional entanglement $\langle E_N^c \rangle$ versus the Coulomb coupling rate $g$ and measurement efficiency $\eta$ for (a)$\sim$(d). (a) and (b), as well as (c) and (d), represent unconditional entanglement under the same feedback strategy and independent feedback strategy, respectively. Specifically, (a) and (c) depict the cases without modulation, while (b) and (d) depict the cases with modulation. Here, we set the charge quantities ratio as $|Q_1|/|Q_2|=3$ and the control effort parameter as $q/\omega_{m}=0.1$. Other parameters are the same as those in Fig.~\ref{fig1}.}
\end{figure}

In Figs.~\ref{fig1}~(a) and (b), we exhibit the  conditional entanglement between two particles as functions of the Coulomb coupling rate $g/\omega_m$ and the detection efficiency $\eta$ for the cases without modulation $\alpha=0$ and with modulation $\alpha=0.2$, respectively. It is evident that, in the case without modulation, the conditional entanglement between the two particles primarily occurs in regions of nearly unstable repulsive interactions ($g/\omega_{m}\simeq-0.25$) and high detection efficiencies. After modulation, conditional entanglement is significantly enhanced, with the parameter requirements being greatly relaxed. This enables the realization of conditional entanglement under low detection efficiency ($\eta \simeq 0.1$) in both repulsive and attractive interaction regimes. To facilitate clearer comparison, we present conditional entanglement as a function of time in Figs.~\ref{fig1}~(c) and (d). It can be observed that the conditional entanglement after modulation increases significantly and oscillates with a period of $1/\Omega$, which is similar to the behavior of quantum entanglement and mechanical squeezing in optomechanical systems with gently modulation \cite{PhysRevA.104.053506, mari2012opto}.

In Fig.~\ref{fig2}, we demonstrate the impact of frequency modulation on unconditional entanglement between particles under different feedback strategies. When the same feedback strategy is chosen, due to the relatively low degree of control over the system at this point, the unconditional entanglement between the two particles is weak. By comparing Fig.~\ref{fig2}(a) with Fig.~\ref{fig2}(b), it can be observed that in the case of frequency modulation, the unconditional entanglement is enhanced, and unconditional entanglement is achieved in both repulsive ($g/\omega_m\simeq-0.2$) and attractive ($g/\omega_m>0.15$) electrostatic interactions. When employing an independent feedback strategy, the advantages of frequency modulation become more pronounced, significantly reducing the threshold of measurement efficiency required to achieve unconditional entanglement. The corresponding results are presented in Figs.~\ref{fig2}(c) and (d). To provide a clearer description, we present the evolution of unconditional entanglement over time in Figs.~\ref{fig2}(e) and (f). It can be observed that regardless of whether the Coulomb interaction manifests as repulsive or attractive, the modulated unconditional entanglement outperforms the unmodulated case.

\begin{figure}\centering
\includegraphics[height=8cm,width=8cm]{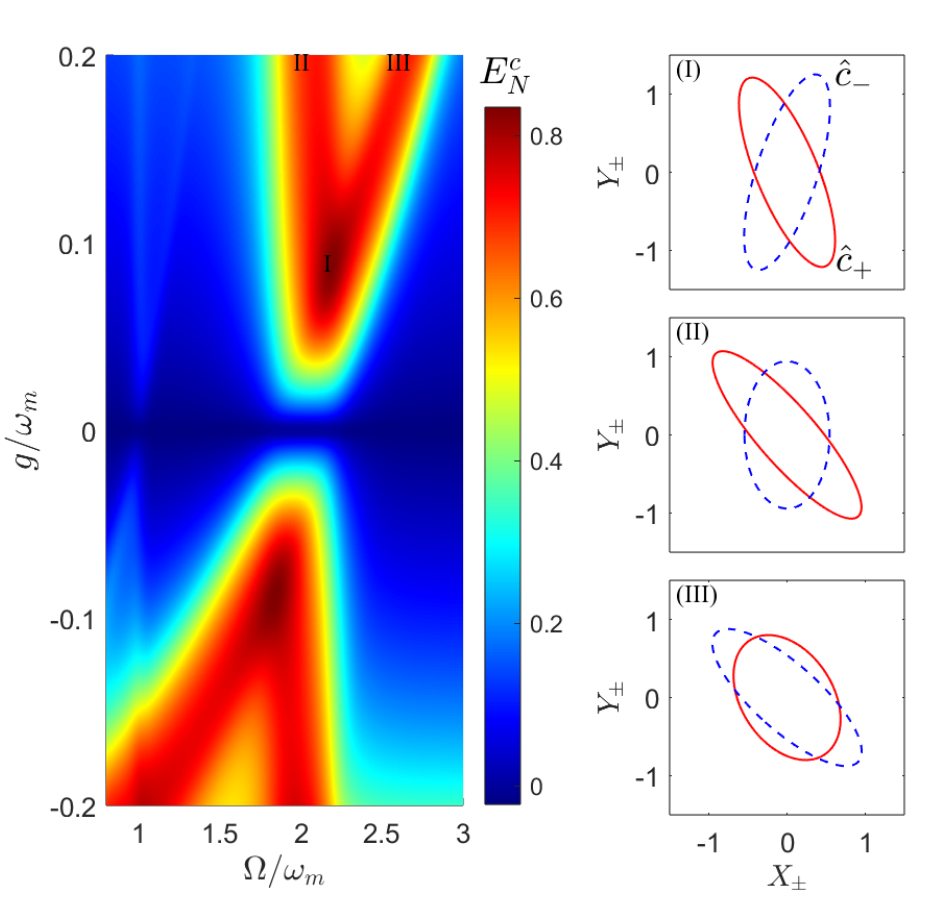}
\caption{\label{fig3}Time-averaged conditional logarithmic negativity $\langle E_N^{\hat c} \rangle$ (averaged over one mechanical period) versus the modulation frequency $\Omega/\omega_m$ and the Coulomb coupling rate $g/\omega_m$ (left). The right figure window displays the uncertainty ellipses of two normal modes $\hat c_+$ (red solid line) and $\hat c_-$ (blue dashed line) for different parameter combinations \{$g/\omega_m$, $\Omega/\omega_m$\}=\{0.86, 2.18\}, \{0.2, 2\}, \{0.2, 2.6\}, labeled (I-III) on the color plot on the left. Other parameters are the same as those in Fig.~\ref{fig1}.}
\end{figure} 
To better analyze the formation of entanglement, we introduce two new normal modes through a beam splitter transformation 
\begin{align}
{\hat{x}_{\pm}} & = \frac{1}{\sqrt{2}}\left( {\hat{x}_{1}}\pm{\hat{x}_{2}} \right), \nonumber  \\
{\hat{p}_{\pm}} & = \frac{1}{\sqrt{2}}\left( {\hat{p}_{1}}\pm{\hat{p}_{2}} \right).
\end{align}
It is worth mentioning that if there exists different quantum squeezing in direction or magnitude between two normal modes, it will lead to the entanglement of two particles. Additionally, the introduction of normal modes can decouple the system dynamics, allowing the system's Hamiltonian Eq.~\ref{hmt1} to be rewritten as
\begin{align}
   \hat H=&\frac{{{\omega }_{m}}}{2}\hat p_{+}^{2}+\frac{{{\omega }_{m}}}{2}{{\left[ 1+\alpha \cos \left( \Omega t \right) \right]}^{2}}\hat x_{+}^{2} \nonumber \\ 
 & +\frac{{{\omega }_{m}}}{2}\hat p_{-}^{2}+\frac{{{\omega }_{m}}}{2}{{\left[ 1+\alpha \cos \left( \Omega t \right) \right]}^{2}}\hat x_{-}^{2}+2g\hat x_{-}^{2}.
 \label{Hpm}
\end{align}
It can be seen that different modulation frequencies will affect the quantum behavior of the two normal modes. Specifically, under weak coupling ($g \ll \omega_m$) and weak modulation ($\alpha \ll 1$), by performing a unitary transformation $U(t)=\text{exp}[-i\frac{\Omega}{2}(\hat x_+^2 + \hat p_+^2 + \hat x_-^2 + \hat p_-^2)t]$ and after applying the rotating-wave approximation, we can obtain an effective Hamiltonian
\begin{align}
  \hat H_{\mathrm{eff} } =&\Delta_+ \hat c_{+}^{\dagger }{\hat{c}_{+}}+\frac{{{\omega }_{m}}\alpha }{4}\left( {\hat{c}_{+}}{\hat{c}_{+}}+\hat c_{+}^{\dagger }\hat c_{+}^{\dagger } \right) \nonumber \\ 
 & + \Delta_- \hat c_{-}^{\dagger }{\hat{c}_{-}}+\frac{{{\omega }_{m}}\alpha }{4}\left( {\hat{c}_{-}}{\hat{c}_{-}}+\hat c_{-}^{\dagger }\hat c_{-}^{\dagger } \right),
 \label{hmt2}
\end{align}
where $\hat c_{\pm} = (\hat x_\pm + i\hat p_\pm)/\sqrt 2$ represents the annihilation operator for the two normal modes, $\Delta_+ = {{\omega }_{m}}+\frac{{{\omega }_{m}}{{\alpha }^{2}}}{4}-\frac{\Omega }{2}$ and $\Delta_- = \Delta_+ + 2g$ denote the effective detunings. It can be observed from Eq.~\ref{hmt2} that there are two resonances, namely $\Omega \simeq 2\omega_m$ and $\Omega \simeq 2\omega_m + 4g$, which are respectively associated with the optimal squeezing of modes $\hat c_+$ and $\hat c_-$ \cite{mari2012opto}. 
\begin{figure}\centering
\includegraphics[height=9cm,width=8cm]{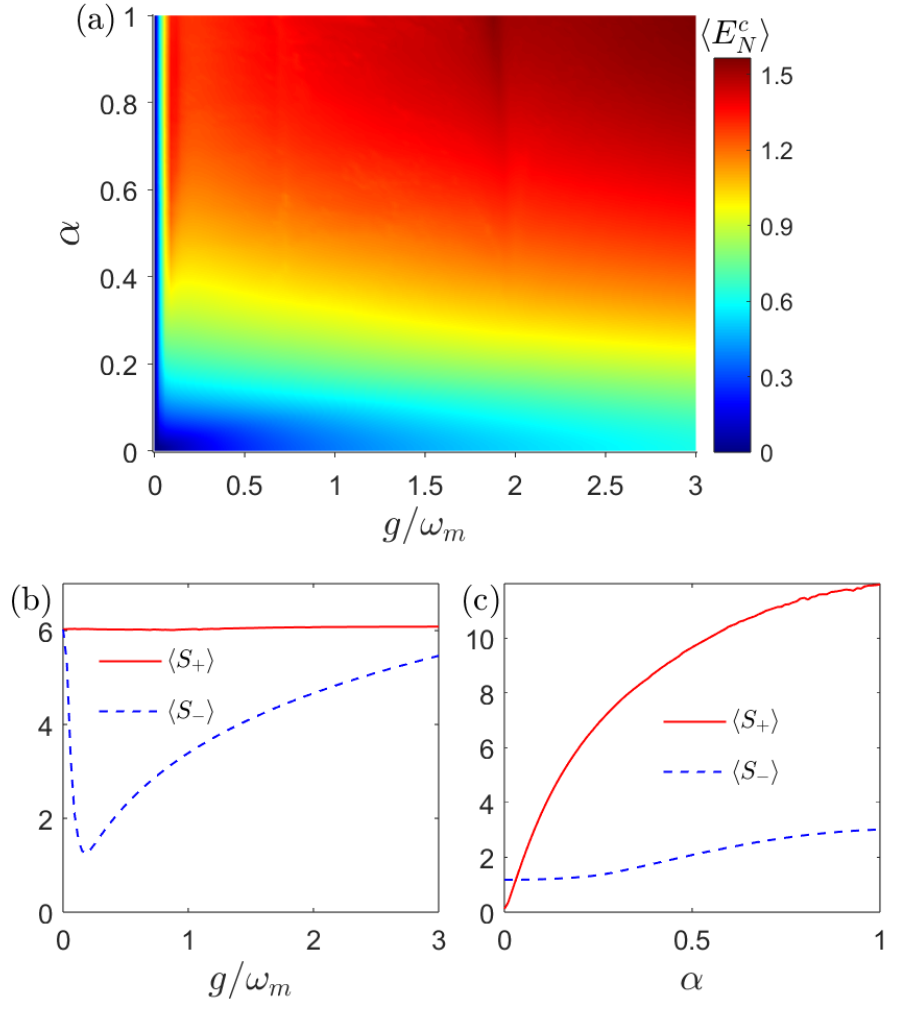}
\caption{\label{fig4}(a) Time-averaged conditional logarithmic negativity $\langle E_N^c \rangle$ versus the Coulomb coupling rate $g/\omega_m$ and the modulation depth $\alpha$. (b) The variation of the squeezing degrees $S_{\pm}$ of the two normal modes $\hat c_{\pm}$ with respect to the Coulomb coupling rate at a modulation depth of $\alpha = 0.2$, and the variation of the squeezing degrees $S_{\pm}$ with respect to the modulation depth at a Coulomb coupling of $g/\omega_m=0.2$. Other parameters are the same as those in Fig.~\ref{fig1}.}
\end{figure} 

To verify our analysis, we exhibit in Fig.~\ref{fig3} the conditional entanglement between two masses as function of the modulation frequency $\Omega/\omega_m$ and the coupling rate $g/\omega_m$. It can be observed that the distribution of conditional entanglement exhibits a distinct double ``V" shape, reaching a peak at the root of the ``V". At this point, both normal modes exhibit significant quantum squeezing, and the squeezing directions are different. In regions away from the root, the conditional entanglement shows a clear bimodal distribution, with one peak concentrated at $\Omega/\omega_m \simeq 2$, where mode $\hat c_+$ exhibits substantial squeezing. The position of the other peak is related to the type of Coulomb interaction: for attractive interactions $(g>0)$, the peak appears at $\Omega/\omega_m>2$, while for repulsive interactions $(g<0)$, the peak appears at $\Omega/\omega_m<2$. In these cases, mode $\hat c_-$ exhibits significant squeezing. The uncertainty ellipses shown in panels (I-III) further confirm the above situation. It is worth noting that homodyne measurement is crucial for the stability of the system. In the absence of homodyne measurement ($\eta=0$), the system reduces to a purely linear one. Under the resonance condition at twice the mechanical frequency ($\Omega=2\omega_m$), the stability condition for the system is given by $\omega_m \alpha/2 < \gamma$, which cannot be satisfied in the regime of mechanical high-quality  factor.

The conditional covariance matrix $\Sigma_c$ associated with the quadrature operators $\hat x_{\pm}$ and $\hat p_{\pm}$ for the two normal modes can be expressed as 
\begin{align}
\left[ \begin{matrix}
   {{\Sigma }_{+}} & \bm 0  \\
   \bm 0 & {{\Sigma }_{-}}  \\
\end{matrix} \right] & = T{{V}_{c}}T,
\end{align}
where ${\Sigma }_{\pm }$ represents the respective two-dimensional covariance matrices corresponding to the two normal modes $\hat c_{\pm}$ , and the symmetric transformation matrix $T$ is give by 
\begin{align}
T=\frac{1}{\sqrt{2}}\left[ \begin{matrix}
   1 & 0 & 1 & 0  \\
   0 & 1 & 0 & 1  \\
   1 & 0 & -1 & 0  \\
   0 & 1 & 0 & -1  \\
\end{matrix} \right].
\end{align}
The squeezing degrees (in units of $\text{dB}$) of the two normal modes are defined as 
\begin{align}
S_{\pm} = -10 \mathrm{log}(2\sigma_{\pm}),
\end{align}
where $\sigma_{\pm}=\mathrm{min}\{\mathrm{eig}[\Sigma_{\pm}]\}$ is the minimum eigenvalue of the two-dimensional covariance matrix $\Sigma_{\pm}$. 

In Fig.~\ref{fig4} (a), we present the evolution of the time-averaged conditional entanglement with respect to the Coulomb coupling rate $g/\omega_m$ and the modulation depth $\alpha$. It is clearly evident that both increasing the modulation depth and the Coulomb coupling rate can enhance the conditional entanglement. However, the increase in modulation depth has a more pronounced effect on entanglement enhancement. This is because, after entanglement is generated, an increase in Coulomb coupling only enhances the squeezing of the normal mode $\hat c_-$, whereas an increase in modulation depth can enhance the squeezing of both normal modes and significantly boost the squeezing amplitude of the normal mode $\hat c_+$. This can be derived from the Hamiltonian Eq.~\ref{hmt2} and is illustrated in Figs.~\ref{fig4}(b) and (c). In addition, strong quantum entanglement ($\langle E_N^c \rangle > \ln 2 \sim 0.69$) can be readily achieved by enhancing the modulation depth, which is not feasible in traditional optomechanical-mechanical entanglement schemes due to stability constraints \cite{PhysRevA.95.053842}. Of course, since the generation of inter-particle entanglement requires a difference in the quantum squeezing of the two normal modes, the presence of Coulomb interaction is crucial for the creation of entanglement.

The computational parameters we employed align with recent experimental conditions \cite{magrini2021real, tebbenjohanns2021quantum, delic2020cooling}. Silica particles with a radius of $R=50~\text{nm}$, a density of $\rho=1850~\text{kg}/\text{m}^3$, and a relative permittivity of $\epsilon_r=2.1$ were trapped by a Gaussian beam featuring a waist radius of $W_t=932~ \text{nm}$ and a wavelength of $\lambda=1550~\text{nm}$. The amplitude at the beam waist was $E=7.4\times 10^{6}~\text{N/C}$ (with a corresponding power of $P=100~\text{mW}$), resulting in a trap frequency of $\omega_m=29.6~\text{kHz}$ and a quantum back-action-induced decoherence rate of $\Gamma_{ba}/\omega_m = 0.053$ \cite{PhysRevA.100.013805}. At room temperature and under one standard atmospheric pressure, one can obtain $\gamma/\omega_m=1.4\times 10^{-11}$ and $\Gamma_\text{th}/\omega_m\simeq2.5\times 10^{-3}$. When an independent feedback strategy is employed, two particles can carry equal charges $Q_{1,2}=30 \text e$, with an inter-particle distance of $d=3\mu \text m$. This leads to a Coulomb coupling rate of $g/\omega_m=0.2$.
\section{CONCLUSION}
In conclusion, we have studied the quantum entanglement between levitated particles coupled via Coulomb interaction by combining frequency modulation with the optimal quantum control strategy. We have found that through frequency modulation, the quantum entanglement between the two particles can be significantly enhanced, and the parameter range for generating entanglement can be substantially broadened. Regarding conditional entanglement, our scheme can achieve strong conditional entanglement $\langle E_N^c \rangle > \ln 2$ under relatively large modulation depth and high measurement efficiency. For unconditional entanglement, by employing independent feedback measurements, it can be realized under weak coupling ($|g|/\omega_m \simeq 0.1$) and low measurement efficiency ($\eta \simeq 0.5$). Finally, we analyzed the mechanism of entanglement generation between particles by introducing normal modes. Frequency modulation induces strong quantum squeezing in the two normal modes, while the Coulomb interaction causes a discrepancy in the squeezing of these two modes, ultimately leading to entanglement between the two particles. Our scheme integrates non-equilibrium dynamics with linear quadratic gaussian technology, significantly enhancing both conditional and unconditional entanglement between levitated particles, thereby providing effective quantum resources for the study of macroscopic quantum effects. 

\textit{Note added}: When completing the paper, we became aware of the brief discussion on the modulation of the trapping frequency had been added in the latest version of \cite{2408.06251}.

\section*{acknowledgment}
This work is supported by the National Natural Science Foundation of China (Nos. 12174140 and 12274274).

\bibliography{apssamp}

\end{document}